\documentclass[runningheads]{svmult}

\usepackage{makeidx}   % allows index generation
\usepackage{graphicx}  % standard LaTeX graphics tool
\usepackage{subeqnar}  % subnumbers individual equations
\usepackage{multicol}  % used for the two-column index
\usepackage{physprbb}  % modified textarea for proceedings,
\makeindex             % used for the subject index
\newcommand{\lmin}{L_{\rm min}}
\newcommand{\lmax}{L_{\rm max}}

\begin{document}
\title*{Determining the Gamma-Ray Burst Rate as a Function of Redshift}
\toctitle{Determining the Gamma-Ray Burst Rate as a Function of Redshift}
\titlerunning{Gamma-Ray Burst Rate as a Function of Redshift}
\author{Nevin Weinberg\inst{1}
\and Carlo Graziani\inst{1}
\and Donald Q. Lamb\inst{1}
\and Daniel E. Reichart\inst{2}
}
\authorrunning{Nevin Weinberg et al.}
\institute{Department of Astronomy \& Astrophysics, University of
Chicago, 5640 South Ellis Avenue, Chicago, IL 60637
\and Department of Astronomy, California Institute of Technology,
Mail Code 105-24, 1201 East California Boulevard, Pasadena, CA 91125}

\maketitle              % typesets the title of the contribution

\begin{abstract}
We exploit the 14 gamma-ray bursts (GRBs) with known redshifts $z$ and
the 7 GRBs for which there are constraints on $z$ to determine the GRB
rate $R_{\rm GRB}(z)$, using a method based on Bayesian inference.  We
find that, despite the qualitative differences between the observed GRB
rate and estimates of the SFR in the universe, current data are
consistent with $R_{\rm GRB}(z)$ being proportional to the SFR.
\end{abstract}

\section{Introduction} 

There is increasing evidence that GRBs are due to the collapse of
massive stars (see, e.g., [1] for a discussion of this evidence).  If
GRBs are indeed related to the collapse of massive stars, one expects
the GRB rate to be roughly proportional to the SFR.  However, the
observed redshift distribution of GRBs differs noticeably from that of
the SFR:  the observed GRB redshift distribution peaks at $z \approx 1$
and few bursts are observed beyond $z\sim 1.5$, while the SFR peaks at
$z \approx 2$ and 10-40\% of stars are thought to form beyond $z=5$
(see, e.g., [2,3,]).

However, observational selection effects play an important role in
determining the observed redshift distribution of GRBs.  The important
question is therefore whether or not the discrepancy between the
observed GRB redshift distribution and the redshift dependence of the
SFR is entirely due to selection effects; i.e., is the GRB rate roughly
proportional to the SFR after taking observational selection effects
into account?  We address this question in this paper.

\begin{figure}
\vskip -8pt
\begin{minipage}[t]{2.435truein}
\mbox{}\\
\includegraphics[width=2.435truein,clip=]{fig1.ps}
\caption{Comparison of the best-fit models to the differential
distribution of peak fluxes of BATSE, IPN and BeppoSAX bursts, and the
cumulative peak flux distributions for the three experiments.}
\end{minipage}
\hfill
\begin{minipage}[t]{2.27truein}
\mbox{}\\
\includegraphics[width=2.27truein,clip=]{fig2.ps}
\caption{Efficiency $\epsilon (P)$ with which BATSE, IPN and BeppoSAX 
detect GRBs as given by the best-fit models to the differential
distribution of peak fluxes of BATSE, IPN and BeppoSAX bursts.}
\end{minipage}
\vskip -10pt
\end{figure}

\section{Method}

We adopt a Bayesian approach.  We calculate the likelihood of the data
given the model, and convert it to a posterior distribution on the
model parameters.  We assume a very general model for the GRB rate, and
a power-law model for the intrinsic GRB photon luminosity distribution
(we assume that the amplitude and the power-law index of the photon
luminosity distribution does not evolve; we relax this assumption in
future work).  We determine the efficiency with which BeppoSAX and the
IPN detect GRBs as a function of peak photon flux $P$ by comparing the
BeppoSAX and IPN peak photon flux distributions to that of BATSE.  We
fit the model jointly to the peak fluxes and redshifts of the 14 GRBs
with known $z$, and the 7 GRBs for which there are constraints on $z$. 
%\subsection{Model}

We write the rate of GRBs that occur per unit redshift and luminosity
as
\vskip -2pt
\begin{equation}
{{dN}/{dz\,dL_N}}=\rho(z)f(L_N) \; ,
\end{equation}
\vskip -5pt
\noindent
where
\begin{equation}
\rho(z)=R_{\rm GRB}(z;P,Q) \times (1+z)^{-1}\times 4\pi r(z)^2 (dr/dz)
\end{equation}
is the rate of GRBs that occur at redshift $z$,
\vskip -2pt
\begin{equation}
R_{\rm GRB}(z;P,Q)=\left[{{t(z)} \over {t(0)}}\right]^P
\exp\left[Q\left(1-{{t(z)} \over {t(0)}}\right)\right]
\end{equation}
is the rate of GRBs that occur at redshift $z$ per unit comoving volume
(see \cite{rr99}), $t(z)$ is the elapsed time since the Big Bang, and
\vskip -2pt
\begin{equation}
f(L_N)=L_N^{-\beta} \times \Theta(L_N-\lmin) \Theta(\lmax-L_N)
\end{equation}
is the intrinsic photon luminosity distribution of GRBs.  Thus the
model has five parameters:  $P$, $Q$, $\beta$, $\lmin$, and $\lmax$.

\begin{figure}
\begin{minipage}[t]{2.3truein}
\mbox{}\\
\includegraphics[width=2.3truein,clip=]{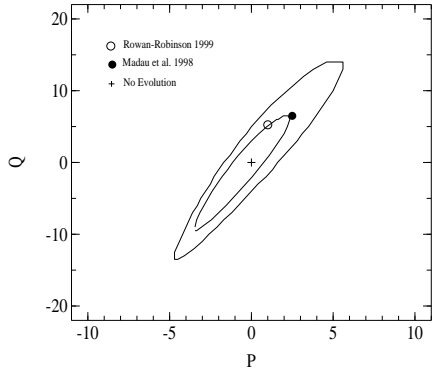}
\end{minipage}
\hfill
\begin{minipage}[t]{2.2truein}
\mbox{}\\
\vskip -17pt
\caption{Probability contours in the (P,Q)-plane for the GRB rate
parameters ($P,Q$) found from fitting to the 14 GRBs with known $z$ and
the 7 GRBs with constraints on $z$.  The solid curves correspond to the
68\% and 95\% probability contours.  Also shown are the (P,Q)-values
corresponding to no space density evolution (+), the Madau et al. SFR
[4], and the Rowan-Robinson phenomenological model fit to IR, optical
and UV data [5].}
\end{minipage} 
\vskip -15pt
\end{figure}

%\subsection{Detection Efficiency}

We write the efficiency with which GRBs with known redshifts are found
as $\epsilon(z,P)=\epsilon_z(z) \times \epsilon_{\rm ST}
\epsilon_P(P)$, where $\epsilon_z(z)$ is the efficiency with which the
redshifts of GRBs are determined from optical observations once they
are detected by a $\gamma$-ray burst instrument.  We take
$\epsilon_z(z) = 1$ for GRBs whose redshifts were determined by
detection of an absorption-line system in the optical afterglow of the
burst and $\epsilon_z(z)=\Theta(1-z) + \Theta(z-2.5)$ for GRBs whose
redshifts were determined by measuring emission lines in the spectra of
the host galaxy.  The latter expression accounts qualitatively for the
difficulty in measuring redshifts when the $H_\alpha$ and O[II]
emission lines from host galaxies do not lie in the visible
spectrum.  The quantity $\epsilon_{\rm ST}$ is the ``stereo-temporal''
efficiency that accounts for limitations of exposure in time and solid
angle and $\epsilon_P(P)$ is the efficiency with which BATSE, the IPN and BeppoSAX detect
GRBs as a function of peak flux $P$.  
%We determine $P_\star$ and $\Delta P$ for BeppoSAX and the IPN by
%comparing their peak flux distributions to BATSE's.   
%First we fit the empirical model
%\begin{equation}
%\frac{d\nb}{dP}\times\epsilon_P(P)=
%\frac{(P/P_s)^a}{1+(P/P_s)^{a+5/2}}\times\epsilon_P(P)
%\end{equation}
%to the BATSE data above the threshold.  Then we fit the model
%\begin{equation}
%\frac{dN}{dP}=\epsilon_P(P)\times\est\times\frac{d\nb}{dP}
%\end{equation}
%to the BeppoSAX and to the IPN data separately.  
Figure 1 compare the best-fit models of $\epsilon_P$ and the cumulative
peak flux distributions of BATSE, the IPN, and BeppoSAX, respectively. 
Figure 2 shows the best-fit $\epsilon_P$ for each of the three
experiments.

%\subsection{Likelihood Function}

The likelihood function is then given by
\vskip -6pt
\begin{equation}
{\cal L}=\exp\left\{-\int dz\,dP\,\mu(z,P)\epsilon(z,P)\right\}
\prod_{i=1}^N \mu(z_i,P_i) \; ,
\end{equation}
\vskip -4pt
\noindent
where
\vskip -12pt
\begin{eqnarray}
\mu(z,P)&=&\int_0^\infty dL_N\,\rho(z)f(L_N)
\delta\left(P-{{L_N} \over {4\pi r(z)^2(1+z)^\alpha}}\right)\nonumber\\
&=&\rho(z)\times f\left(4\pi r(z)^2(1+z)^\alpha P\right)
\times 4\pi r(z)^2(1+z)^\alpha
\end{eqnarray}
is the expected number of events observed within $dz\,dP$ of $(z,P)$.
The quantity $\alpha$ is the burst spectral index, which we set equal
to one in this work.  By an application of Bayes' Theorem, we now
regard ${\cal L}$ as an (unnormalized) probability distribution on the
model parameters.

\section{Results}

Figure 3 shows 68\% and 95\% probability contours for the GRB rate
parameters $P$ and $Q$.  Also shown on the plots are the best-fit SFR
models of Madau et al. [4] and Rowan-Robinson [5].  The SFR models lie
at about a 68\% excursion from the best-fit GRB rate model.  Thus
we find that, despite the qualitative differences that exist between
the observed GRB rate and estimates of the SFR in the universe, current
data are consistent with the actual GRB rate being approximately
proportional to the SFR.

\end{document}